# $Fe_{1+y}Te_xSe_{1-x}$: a delicate and tunable Majorana material

Fazhi Yang (杨发枝)[1,2]*, Giao Ngoc Phan[1,3]*†, Renjie Zhang (张任杰)[1,2], Jin Zhao (赵金)[1,2], Jiajun Li (李佳俊)[1,2], Zouyouwei Lu (鲁邹有为)[1,2], John Schneeloch[4], Ruidan Zhong (钟瑞丹)[4,6], Mingwei Ma (马明伟)[1,5], Genda Gu (顾根大)[4], Xiaoli Dong (董晓莉)[1,2,5], Tian Qian (钱天)[1,5], Hong Ding (丁洪)[1,3,6]‡

[1]*Beijing National Laboratory for Condensed Matter Physics and Institute of Physics, Chinese Academy of Sciences, Beijing 100190, China*
[2]*School of Physics, University of Chinese Academy of Sciences, Beijing 100190, China*
[3]*CAS Center for Excellence in Topological Quantum Computation, University of Chinese Academy of Sciences, Beijing 100190, China*
[4]*Condensed Matter Physics and Materials Science Department, Brookhaven National Laboratory, Upton, NY, USA*
[5]*Songshan Lake Materials Laboratory, Dongguan 523808, China*
[6]*Tsung-Dao Lee Institute, Shanghai Jiao Tong University, Shanghai 201210, China*

Corresponding authors. Email: dingh@sjtu.edu.cn and gphan@iphy.ac.cn

We report the observation for the $p_z$ electron band and the band inversion in $Fe_{1+y}Te_xSe_{1-x}$ with angle-resolved photoemission spectroscopy. Furthermore, we found that excess Fe (y>0) inhibits the topological band inversion in $Fe_{1+y}Te_xSe_{1-x}$, which explains the absence of Majorana zero modes in previous reports for $Fe_{1+y}Te_xSe_{1-x}$ with excess Fe. Based on our analysis of different amounts of Te doping and excess Fe, we propose a delicate topological phase in this material. Thanks to this delicate phase, one may be able to tune the topological transition via applying lattice strain or carrier doping.

Iron-based superconductors (FeSCs) become a promising platform for long-sought-after topological quantum computation thanks to recent discoveries of topological surface states and Majorana zero modes (MZMs) in them [1-7]. The band inversion between the electron band of the chalcogen $p_z$ orbital character and the hole band of the Fe $d_{xz}$ character along the $\Gamma-Z$ direction is crucial to realize the topological non-trivial surface states in FeSCs [8,9]. Fe(Te, Se) (or more precisely $Fe_{1+y}Te_xSe_{1-x}$) has the simplest structure among FeSCs. As doping Te into FeSe, the interlayer $p$-$p$ coupling enhancement causes stronger $k_z$ dispersion of the $p_z$ band, bringing the $p_z$ band close to the $d_{xz}$ band. When the $p_z$ and $d_{xz}$ bands cross each other along the $\Gamma-Z$ direction, the topological band inversion occurs, and a spin-orbit coupling (SOC) gap is opened between those bands. However, direct experimental evidence of the $p_z$ band and the band inversion in Fe(Te, Se) is still a missing puzzle piece [10,11]. In Fe(Te, Se), the MZMs were observed only in a portion of the vortices induced by the external magnetic field [2,12] or even completely absent in the samples with excess Fe [13]. The former may be caused by hybridization induced by the vortex lattice [14,15] and the chemical potential fluctuations due to the strong inhomogeneity in Fe(Te, Se) [16-18]. However, the absence of MZMs at the vortices in the excess-Fe samples $Fe_{1+y}Te_xSe_{1-x}$ has been a long-lasting puzzle.

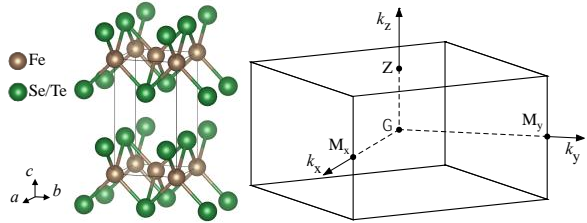

**Fig. 1.** (a) The crystal structure of Fe(Te, Se). (b) The corresponding Brillouin zone.

In this letter, we present a direct observation of the electron $p_z$ band in Fe(Te, Se) and the excess-Fe effect on the band inversion in $Fe_{1+y}Te_{0.7}Se_{0.3}$ with angle-resolved photoemission spectroscopy (ARPES). The direct observations of the $p_z$ electron band and the band inversion confirm the MZM nature of scanning-tunneling-microscopy/spectroscopy (STM/S) results reported before. The observed excess-Fe effect in $Fe_{1+y}Te_{0.7}Se_{0.3}$ gives a key insight into the absence of MZMs at the vortices in the excess-Fe samples [13]. Our analyses strongly suggest the delicate topological property in $Fe_{1+y}Te_xSe_{1-x}$, offering a practical platform for studying tunable MZMs.

The crystal structure of Fe(Te, Se) is shown in Fig. 1(a). The excess Fe could exist at Fe(Te, Se) interlayers [19]. High-quality FeSe single crystals were grown by the chemical vapor transport method using a mixture of $AlCl_3$ and KCl as the transport agent [20]. High-quality $FeTe_{0.55}Se_{0.45}$ single crystals, free from excess Fe, were grown using the self-flux method. These samples originate from the same sample batch used in the previous STS report of the vortex MZMs [2]. High-quality $Fe_{1+y}Te_{0.7}Se_{0.3}$ single crystals were grown by the unidirectional solidification method. With the method, an amount of excess Fe exists in as-grown samples $Fe_{1+y}Te_{0.7}Se_{0.3}$, and the excess Fe can be removed gradually via post-growth annealing in a sealed-vacuum-quartz tube to become $Fe_{1+k}Te_{0.7}Se_{0.3}$ ($k < y$). High-resolution ARPES measurements were carried out with an R4000 analyzer and a helium discharge lamp. The energy resolution is set at 5 meV, and the sample temperature is set at 30 K for all ARPES measurements.

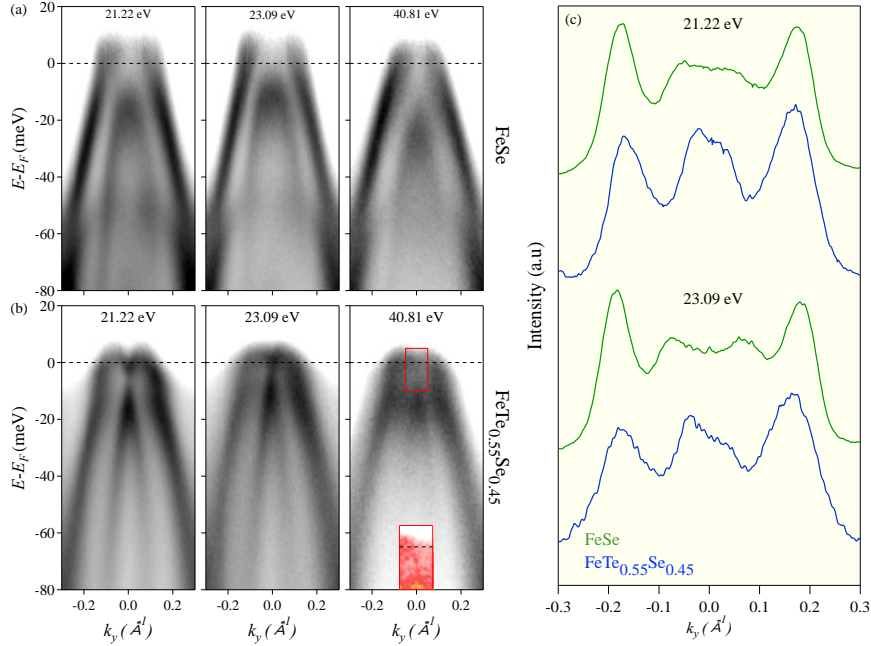

**Fig. 2.** (a) The ARPES intensities divided by the Fermi-Dirac distribution function at $T$ = 30 K of FeSe measured along $\Gamma - M$ with three different photon energies. The $d_{xz}$ lower-hole-band intensity is weaker than that of the $d_{yz}$ upper-hole-band due to the matrix element effect. These two hole-bands exhibit obvious $k_z$ dispersion. (b) Similar to (a) but measured for FeTe$_{0.55}$Se$_{0.45}$. A shallow $p_z$ electron band emerges above the Fermi level with strong $k_z$ dispersion. (c) Comparison of the momentum distribution curves (MDCs) at the energy of 25 meV bellow the Fermi level for FeSe and FeTe$_{0.55}$Se$_{0.45}$ measured with different photon energies. The inner peaks correspond to the $d_{xz}$ band, and the outer peaks correspond to the $d_{yz}$ band. The intensity enhancement of the $d_{xz}$ lower-hole-band top by factor of three indicates the band inversion between the $p_z$ and $d_{xz}$ bands. Inset: A weak linear-dispersion state can be resolved thanks to the absence of the $p_z$ band in its vicinity, which is similar Dirac-cone surface states observed with laser ARPES reported before [1].

Figure 2(a) shows the ARPES intensities divided by the Fermi distribution function of FeSe. The ARPES intensities were measured along the $\Gamma - M$ direction with three different photon energies of 21.22, 23.09, and 40.81 eV corresponding to $k_z$ of 4.53, 4.70, and 6.03 $\pi/c$, respectively. Two hole bands with clear $k_z$ dispersion are observed along the Brillouin-zone center line $\Gamma - Z$ [21,22]. The $d_{xy}$ spectral intensity is too weak to be observed due to its large decoherence caused by strong orbital-selective correlations [23]. Benefited from the partially polarized light source, we can analyze the orbital character of these two hole bands using the matrix element effect. With the difference in the spectral intensity by a factor of three, the upper hole band with stronger spectral intensity and the lower band with weaker spectral intensity are mainly of $d_{yz}$ and $d_{xz}$ orbital characters, respectively. This orbital character is consistent with the experimental results reported for Fe(Te, Se) [22]. In FeTe$_{0.55}$Se$_{0.45}$, similar to FeSe, a set of two-hole bands is also observed. Interestingly, we observed a shallow electron band emerging from above the Fermi level with the photon energies of 21.22 and 23.09 eV corresponding to $k_z$ of 4.97 and 5.15 $\pi/c$, respectively, which are close to the $Z$ point. At the same time, the electron band could not be observed with the photon energy of 40.81 eV corresponding to $k_z$ of 6.61 $\pi/c$, which is between the $\Gamma$ and $Z$ points. The behavior of electron band is consistent with the $p_z$ band predicted by theory and can be assigned to it [8]. Furthermore, the spectral intensity of the lower-hole-band top in FeTe$_{0.55}$Se$_{0.45}$ is enhanced by a factor of three to those in FeSe shown in the comparison of the momentum distribution curves (MDCs) in Fig. 2(c), indicating that the $d_{xz}$ band may have undergone a band inversion with the $p_z$ electron band. A linear-dispersion state can be resolved with the photon energy of 40.81 eV thanks to the absence of the $p_z$ band in its vicinity at $k_z$ away from the $Z$ point. The linear dispersion is similar to Dirac-cone surface states observed with laser ARPES reported before [1]. The emergence of the shallow electron band at the Fermi level in FeTe$_{0.55}$Se$_{0.45}$ with pronounced $k_z$ dispersion confirms the previous theoretical prediction that Te doping brings the $p_z$ band close to $d_{xz}$, promoting the topological band inversion and formation of the topological surface states.

To investigate the effect of Te doping and excess Fe in Fe$_{1+y}$Te$_x$Se$_{1-x}$, we measured Fe$_{1+y}$Te$_{0.7}$Se$_{0.3}$ samples with different amounts of excess Fe. The ARPES intensity of the as-grown Fe$_{1+y}$Te$_{0.7}$Se$_{0.3}$ and annealed Fe$_{1+k}$Te$_{0.7}$Se$_{0.3}$ ($k < y$) are respectively shown in Figs. 3(a) and 3(b). In contrast to the FeTe$_{0.55}$Se$_{0.45}$, the spectral intensity of the lower-hole-band top is much weaker or invisible with the photon energy of 40.81 eV. The weakening of intensity by a factor of three in $d_{xz}$ band top intensity can be seen with the comparison of MDCs and energy distribution curves (EDCs) shown in Figs. 3(c) and 3(d), respectively. The strong and sharp peak of $d_{xz}$ in FeTe$_{0.55}$Se$_{0.45}$ becomes weak and broaden in the as-grown Fe$_{1+y}$Te$_{0.7}$Se$_{0.3}$ and annealed Fe$_{1+k}$Te$_{0.7}$Se$_{0.3}$. This suggests that the band inversion does not occur in Fe$_{1+y}$Te$_{0.7}$Se$_{0.3}$ samples with excess Fe. Therefore in Fe$_{1+y}$Te$_{0.7}$Se$_{0.3}$, both Te-doping and excess-Fe effects are involved simultaneously. To figure out the individual role of each effect on the band structure, we analyze the energy band positions for these observed spectra and plot the results as a function of $k_z$. $d_{xz}$ Band top and $p_z$ band bottom are extracted from the EDCs by two Lorentzian fitting; because the $d_{yz}$ band top is in the unoccupied energy range, the $d_{yz}$ top is determined via parabolic fitting of the $d_{yz}$ band dispersions extracted from the momentum distribution curves (MDCs) by Lorentzian fitting.

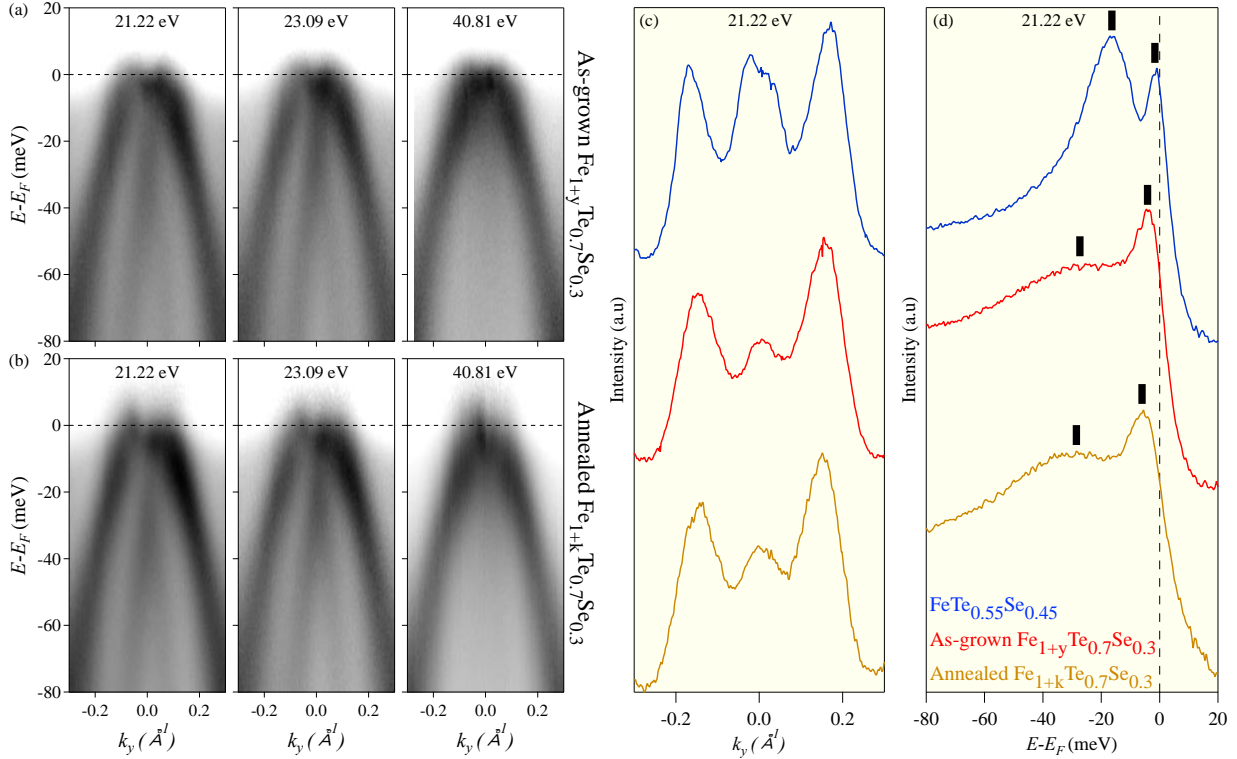

**Fig. 3.** (a) ARPES intensities of as-grown $Fe_{1+y}Te_{0.7}Se_{0.3}$ with different photon energies. (b) Same as (a) but for annealed $Fe_{1+k}Te_{0.7}Se_{0.3}$ ($k < y$). (c) Comparison of MDCs at the energy of 25 meV bellow the Fermi level for $FeTe_{0.55}Se_{0.45}$ in Fig 2(b), as-grown $Fe_{1+y}Te_{0.7}Se_{0.3}$, and annealed $Fe_{1+k}Te_{0.7}Se_{0.3}$ measured with photon energy of 21.22 eV. The inner peaks correspond to the $d_{xz}$ band, and the outer peaks correspond to the $d_{yz}$ band. (d) Similar to (c) but for the energy distribution curves (EDCs) at $k_y = 0$. The bars indicate the peaks extracted from Lolentzian fitting. In contrast to $FeTe_{0.55}Se_{0.45}$, no intensity enhancement of the lower hole-band top implies there is no band inversion in $Fe_{1+y}Te_{0.7}Se_{0.3}$ and $Fe_{1+k}Te_{0.7}Se_{0.3}$. The $p_z$ band bottom shifts downward in the annealed $Fe_{1+k}Te_{0.7}Se_{0.3}$ compared to the as-grown $Fe_{1+y}Te_{0.7}Se_{0.3}$.

Since the $c$-axis increases monotonously to the Te doping level ($x$) shown in Fig. 4(f), the same photon energy probes at different $k_z$ values for each doping level; therefore, we present our analysis of band positions as a function of the relative $k_z$, which is the relative distance to its nearest $\Gamma$ point. The energy positions of the $p_z$ band bottom, and the $d_{yz}$, $d_{xz}$ band tops are plotted as a function of the relative $k_z$ shown in Figs. 4(a-c). $\Delta E$ is defined as the energy gap between the $p_z$ band bottom and the $d_{xz}$ band top, as shown in Fig. 4(d). Increasing the Te doping level, the $d_{yz}$ and $d_{xz}$ bands seem less dispersive, which can be expected from the $c$-axis expansion Fig. 4(f) [24]. Generally speaking, the $c$-axis expansion weakens the $k_z$ dispersion in the band structure. In the vicinity of the $Z$ point, the $d_{xz}$ and $d_{yz}$ bands shift downward with the in-plane-lattice expansion as increasing Te doping level shown in Fig. 4(f), in agreement with the in-plane-lattice-strain effect reported before [25]. A similar behavior is also observed for the $p_z$ electron band. From FeSe to FeTe$_{0.55}$Se$_{0.45}$, the $p_z$ band shifts downward from about 20 meV above the Fermi level [26] to the Fermi level vicinity at $k_z$ around the $Z$ point, and the $p_z$ band slightly shifts downward more as increasing Te doping level to Fe$_{1+y}$Te$_{0.7}$Se$_{0.3}$. It is noteworthy that the energy position change of the $d_{xz}$ band is more than twice that of the $p_z$ band around the $Z$ point from FeTe$_{0.55}$Se$_{0.45}$ to Fe$_{1+y}$Te$_{0.7}$Se$_{0.3}$, enlarging $\Delta E$ in Fe$_{1+y}$Te$_{0.7}$Se$_{0.3}$. The increase in $\Delta E$ obstructs the topological band inversion between $p_z$ and $d_{xz}$ bands.

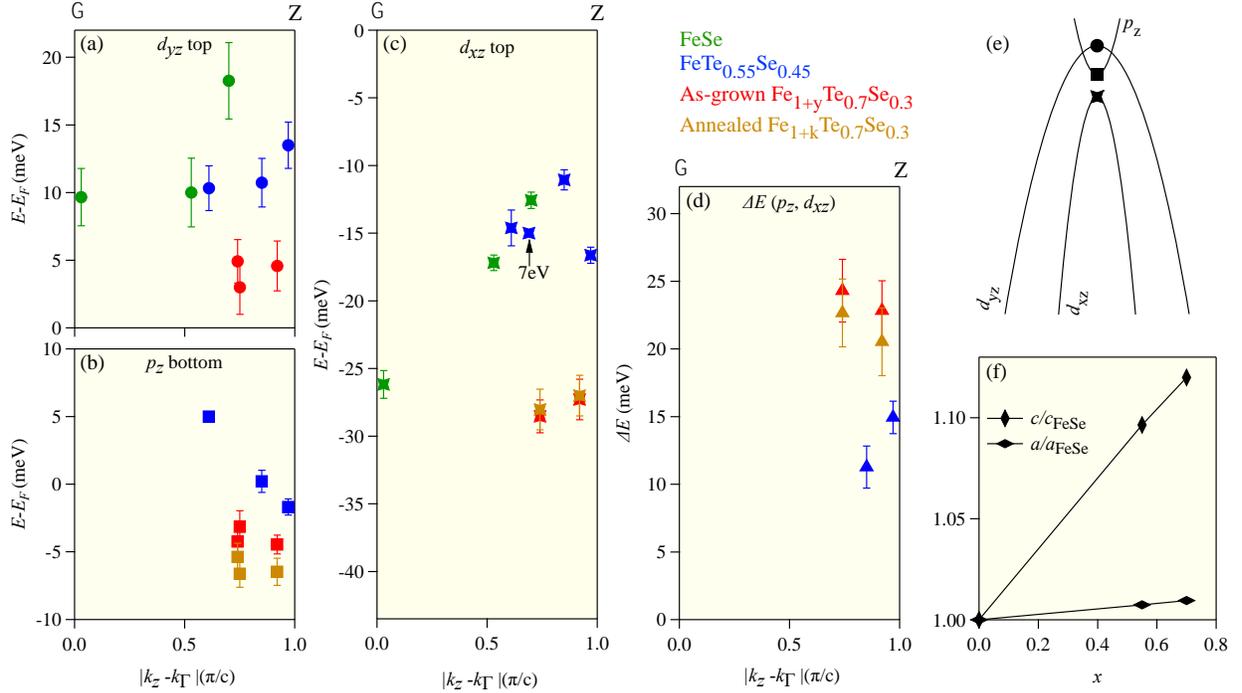

**Fig. 4.** (a)-(c) The energy positions of the $d_{yz}$ band top, the $p_z$ band bottom, and the $d_{xz}$ band top (illustrated in (e)) plotted as a function of the relative $k_z$ distance to its nearest $\Gamma$ point. (d) The energy gap ($\Delta E$) between the $p_z$ band bottom and the $d_{xz}$ band top plotted as a function of the relative $k_z$. (e) Schematic of the band structure in Fe(Te, Se). (f) The ratio of lattice parameters between different Te doping levels ($x$) and FeSe [24].

Furthermore, reduction of excess Fe shows negligible effects on the $d_{yz}$ and $d_{xz}$ hole bands but a visible effect on the $p_z$ electron band. The $p_z$ band shifts downward, resulting in a smaller $\Delta E$, which is only 5 meV larger than that in FeTe$_{0.55}$Se$_{0.45}$. However, no band inversion could be observed for the Fe$_{1+k}$Te$_{0.7}$Se$_{0.3}$. In other words, excess Fe causes the $p_z$ band shifts upward away from the $d_{xz}$ band, obstructing the topological band inversion. For promoting the topological band inversion, annealing for a long time can altogether remove excess Fe in $x = 0.7$, bringing the $p_z$ band close to the $d_{xz}$ band. That is probably why Johnson *et al.* observed the topological band inversion in FeTe$_{0.7}$Se$_{0.3}$ with laser ARPES [27,28]. Surprisingly, the difference of 5 meV in the $\Delta E$ between FeTe$_{0.55}$Se$_{0.45}$ and Fe$_{1+k}$Te$_{0.7}$Se$_{0.3}$ causes a topological phase transition from the topological phase to the non-topological phase. Because of the sensitivity to $\Delta E$, we propose the delicate topological phase in Fe$_{1+y}$Te$_x$Se$_{1-x}$.

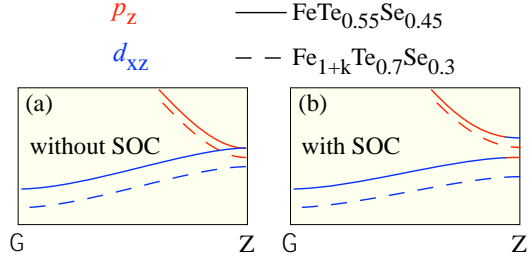

Fig. 5. Schematic of the band structures of different Te doping levels. FeTe$_{0.55}$Se$_{0.45}$ and Fe$_{1+k}$Te$_{0.7}$Se$_{0.3}$ are represented by solid and dashed lines, respectively. The $d_{xz}$ and $p_z$ orbital band are labeled by red and blue colors. (a) Schematic of band structures without SOC. (b) Schematic of band structures with SOC.

Figure 5 illustrates the delicate topological phase in Fe(Te, Se). For FeTe$_{0.55}$Se$_{0.45}$, shown by the solid lines, the $p_z$ band touches with the $d_{xz}$ band around the Z point when SOC is not considered (Fig. 5(a)). When SOC is considered, it opens a SOC gap of about 10 meV between the $p_z$ and $d_{xz}$ bands since there is no rotation symmetry protection, forming the band inversion (Fig. 5(b)) [8]. Without SOC, as the $p_z$ and $d_{xz}$ bands just touch together in the topological phase, any effect that opens a crystal band gap between $p_z$ and $d_{xz}$ would lift the touching. Thus, in Fe$_{1+k}$Te$_{0.7}$Se$_{0.3}$ shown by dashed lines, without SOC, the excess Fe opens a crystal band gap between $p_z$ and $d_{xz}$ bands along the $\Gamma - Z$ direction due to a greater downward shift of the $d_{xz}$ band in comparison with that of the $p_z$ band as shown in Fig. 5(a). When SOC is considered, the crystal band gap combines with the SOC gap, forming a larger $\Delta E$ than that in FeTe$_{0.55}$Se$_{0.45}$ as shown in Fig. 5(b). This larger $\Delta E$ in Fe$_{1+k}$Te$_{0.7}$Se$_{0.3}$ is consistent with our observation. Due to the delicate topological phase, besides the Te doping, the excess Fe is also a key parameter to control topological phase in $Fe_{1+y}Te_xSe_{1-x}$. The observed excess-Fe effect and the delicate topological phase can explain the absence of MZMs in some vortices in $FeTe_{0.55}Se_{0.45}$ or almost all vortices in the excess-Fe samples [2,12,13]. Interestingly, the delicate topological phase in $Fe_{1+y}Te_xSe_{1-x}$ makes itself a candidate for a tunable MZM platform as its $\Delta E$ can be practically manipulated using other parameters such as lattice strain or carrier doping.

In summary, we observed direct experimental evidence of the $p_z$ electron band and the topological band inversion in $Fe_{1+y}Te_xSe_{1-x}$. Our analysis results show that the excess Fe is a crucial parameter for controlling the topological phase in $Fe_{1+y}Te_xSe_{1-x}$, which makes it a candidate for tunable MZM platform by fine-tuning its delicate topological phase.

We thank Zhijun Wang, Lingyuan Kong, and Wenyao Liu for helpful discussions. This work was performed at Beijing National Laboratory for Condensed Matter Physics and Institute of Physics, Chinese Academy of Sciences, Beijing, China.

Supported by the National Natural Science Foundation of China (Grant Nos. 11888101 and U1832202), the Chinese Academy of Sciences (Grant Nos. QYZDB-SSW- SLH043, XDB28000000, and XDB33000000), the K. C. Wong Education Foundation (Grant No. GJTD-2018-01), and the Informatization Plan of Chinese Academy of Sciences (Grant No. CAS-WX2021SF-0102). This work was also supported by the Synergetic Extreme Condition User Facility (SECUF). G.D.G. was supported by US DOE (Grant Nos. DE-SC0010526 and DE-SC0012704).